\documentclass[twocolumn,showpacs,preprintnumbers,amsmath,amssymb]{revtex4}
\usepackage{amsmath,amssymb,graphics,epsfig,subfigure}
\usepackage{color}

\begin{document}
\thispagestyle{empty}

\begin{center}

\title{Characteristic interaction potential of black hole molecules from the microscopic interpretation of Ruppeiner geometry}

\date{\today}
\author{Shao-Wen Wei$^{1,2,3}$\footnote{Corresponding author. E-mail: weishw@lzu.edu.cn},
        Yu-Xiao Liu$^{1,2,3}$\footnote{E-mail: liuyx@lzu.edu.cn}, and
        Robert B. Mann$^{4}$ \footnote{E-mail: rbmann@uwaterloo.ca}}

\affiliation{$^{1}$Lanzhou Center for Theoretical Physics, Key Laboratory of Theoretical Physics of Gansu Province, School of Physical Science and Technology, Lanzhou University, Lanzhou 730000, People's Republic of China,\\
 $^{2}$Institute of Theoretical Physics $\&$ Research Center of Gravitation,
Lanzhou University, Lanzhou 730000, People's Republic of China,\\
 $^{3}$Academy of Plateau Science and Sustainability, Qinghai Normal University, Xining 810016, P. R. China\\
 $^{4}$Department of Physics and Astronomy, University of Waterloo, Waterloo, Ontario, Canada, N2L 3G1}

\begin{abstract}
Ruppeiner geometry has been found to be a novel promising approach to uncover the microstructure of fluid systems and black holes. In this work, combining with the micro model of the Van der Waals fluid, we shall propose a first microscopic interpretation for the empirical observation of Ruppeiner geometry. Then employing the microscopic interpretation, we disclose the potential microstructure for the anti-de Sitter black hole systems. Of particular interest, we obtain the microscopic interaction potentials for the underlying black hole molecules. This significantly strengthens the study towards to the black hole nature from the viewpoint of the thermodynamics.
\end{abstract}


\maketitle
\end{center}

\section{Introduction}

Classically a black hole is a purely absorbing object; it cannot emit particles. However Hawking showed that when quantum mechanical effects are considered near the horizon, black holes can indeed emit particles, with an average energy flux given by \cite{Hawking}
\begin{equation}
 \frac{dE}{dt}=\int_{0}^{\infty}\frac{\Gamma\omega}{e^{\omega\beta}-1}d\omega,
\end{equation}
where $\Gamma$ is the greybody factor of the radiation. This suggests that a black hole emits particles at a temperature $T_{\rm H}=1/\beta$ known as the Hawking temperature, which depends on the black hole mass, charge and other parameters. Furthermore, the corresponding thermodynamic entropy is $S=\mathcal{A}k_{\rm B}c^3/4\hbar G$ with $\mathcal{A}$ the horizon area. The four laws of black hole thermodynamics were subsequently established \cite{Hawking,Bekenstein,Bardeen} and it is now extensively believed that black holes are thermodynamic systems. In contrast to an ordinary fluid system, black hole entropy is proportional to the horizon area of the black hole rather than its volume. This brings new peculiar properties to black holes.

An outstanding challenge is that of understanding the microstructure of a black hole that underlies this thermodynamics. Different theories and models have been proposed, including string theory \cite{Vafa,Maldacena,Callan,Horowitz}, fuzzball models \cite{Lunin,Mathur}, and pierced horizons \cite{Rovelli:1996dv}, each with its own limitations.

Over the past two decades there has been considerable interest in black hole phase transitions in anti-de Sitter space \cite{Chamblin,Chamblin2,Cognola,Kubiznak}. Motivated by the interpretation of the cosmological constant as a thermodynamic pressure \cite{Creighton:1995au,Kastor}, four-dimensional charged AdS black holes were found to exhibit a small-large black hole phase transition, fully analogous to the liquid-gas phase transition of a Van der Waals (VdW) fluid \cite{Kubiznak}. Later studies also showed that this small-large black hole phase transition is ubiquitous, taking place in most black hole systems \cite{Kubiznak:2016qmn}.

It is well known that phase transitions result from the coexistence and competition among different constituents of a physical system. Phase transitions thereby provide insight into the properties of a given system's microstructure. An effective method, Ruppeiner geometry \cite{Ruppeiner}, has been developed to explore this idea. It has been successfully applied to many fluid systems, including VdW fluids, and many interesting microscopic properties have been uncovered, shedding new light on thermodynamic systems. Applying this approach, in recent work \cite{Weiw,Weiwa2,WeiWeiWei} we proposed that the constituent degrees of freedom of black holes have a molecular character. Specifically, we found that under certain circumstances repulsive microstructure interactions could exist for black holes, in strong contrast to VdW fluids \cite{Weiwa2,WeiWeiWei}. However the origin of the repulsive/attractive interactions remains to be solved.

The present work focuses on this significant problem. We first provide a microscopic interpretation of Ruppeiner geometry, and then apply it to black hole systems. We also find the explicit potentials these underlying black hole molecules experience. We take the following units $G=c=k_{B}=l_{P}=1$.

\section{Microscopic interpretation of Ruppeiner geometry}

Ruppeiner observed for various fluid systems that a negative/positive curvature scalar of the geometry was related to a respective attractive/repulsive interaction of the fluid molecules \cite{Ruppeiner}. Employing this property, Ruppeiner geometry has been used to examine the microstructure of some complex fluid systems and black holes. Why this empirical observation works is still a mystery. Since Ruppeiner geometry is closely related to phase transitions, we begin by providing a microscopic interpretation of Ruppeiner geometry for the VdW fluid model.

\begin{figure}
\includegraphics[width=7cm]{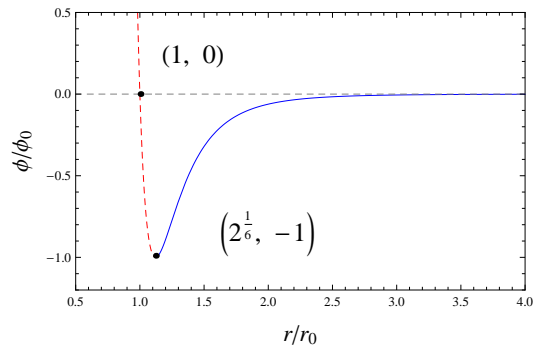}
\caption{The Lennard-Jones potential. The red dashed curve and blue solid curve are for the short-range repulsive interaction and longer-range attractive interaction, respectively.}\label{LJpotential}
\end{figure}

Comparing to an ideal fluid, one needs to take into account the size of and interaction between the molecules for the VdW fluid, which are described by the parameters $b$ and $a$, respectively. The equation of state reads
\begin{equation}
 P=\frac{NT}{V-Nb}-\frac{N^2a}{V^2}=\frac{T}{v-b}-\frac{a}{v^2},\label{eos}
\end{equation}
where $P$, $T$, $V$, and $v$ are the pressure, temperature, total volume, and specific volume of the system, and $N$ corresponds to the number of the molecules contained in the fluid. The equation of state shows a first-order phase transition, and admits a second order phase transition point at the critical point ($P_{\rm c}$, $T_{\rm c}$, $v_{\rm c}$)=($\frac{a}{27b^2}$, $\frac{8a}{27b}$, $3b$). Such phenomena can be understood from a microscopic viewpoint. As is well known, the interaction between two fluid molecules is given by the Lennard-Jones potential \cite{Jones}
\begin{equation}
 \phi=4\phi_{0}\left[\left(\frac{r_{0}}{r}\right)^{12}-\left(\frac{r_{0}}{r}\right)^{6}\right]
\end{equation}
illustrated in Fig. \ref{LJpotential}. The red dashed curve and blue solid curve are for the short-range repulsive interaction and longer-range attractive interaction, respectively. At $r=r_{0}$, we have $\phi$=0, with the minimum of the potential $-\phi_{0}$ occurring at $r_{\rm min}=2^{\frac{1}{6}}r_{0}$. A VdW fluid is described by a ``hard-core" model, meaning that the interatomic distance between molecules cannot be smaller than a molecular diameter $d=r_{\rm min}$. Hence there is a cutoff at the location of the well, and the short-range repulsive interaction denoted by the red dashed curve is excluded. Furthermore, taking the mean field approximation, the characteristic parameters $a$ and $b$ in the equation of state (\ref{eos}) have a microscopic interpretation \cite{Johnstona}
\begin{equation}
 a=\frac{20\pi r_{0}^{3}\phi_{0}}{9\sqrt{2}},\quad b=\sqrt{2}r_{0}^3.\label{abea}
\end{equation}
where $a > 0$, indicating the attractive interaction dominates the fluid microstructure.

We now clearly address two issues for a VdW fluid that are often neglected. The first one is the volume $v\geq b$. When $v=b$, it means the molecules completely occupy the space of the system. No spatial degrees of freedom are left, and thus one may conjecture there is no interaction between these molecules. Furthermore, we can expect that when this condition is violated repulsive interactions will be present. Second, when $T<\frac{27}{32}T_{\rm c}=\frac{a}{4b}$, the pressure is negative for a given range of $v$, which may lead to the repulsive interaction.

Making use of fluctuation theory, Ruppeiner geometry has been used to great success in revealing the intrinsic microstructure of black holes. In particular, the normalized Ruppeiner curvature scalar was introduced and shown for black holes to have universal behaviour near a critical point \cite{Weiwa2,WeiWeiWei}. Here we consider the relation between the interactions and the normalized curvature scalar, which for a VdW fluid is given by
\begin{eqnarray}
 R_{\rm N}=-\frac{2 a (v-b) \left(a b+P
   v^3\right)}{\left(2 a b-a v+P
   v^3\right)^2}\; .\label{rrn}
\end{eqnarray}
Obviously, the denominator of $R_{\rm N}$ is non-negative, vanishing when the system approaches the critical point, which leads to $R_{\rm N}\sim (T_{\rm c}-T)^{-2}$. Here we are concerned with the numerator of $R_{\rm N}$. Noting that the parameters $a$ and $b$ are positive, we see that $R_{\rm N} < 0$ when $v\geq b$. This implies an attractive interaction, which is consistent with the microscopic potential model of the VdW fluid.

We can gain further insight by relaxing the requirements on these parameters.
First, we see that $R_{\rm N}$ vanishes when $a=0$. Since $a$ is a measure of molecular interaction, vanishing $a$ indicates there is no interaction, consistent with empirical observations of Ruppeiner geometry. As expected, from Fig. \ref{LJpotential} and Eq. (\ref{abea}), positive/negative $a$ corresponds to a well/barrier, indicating attractive/repulsive interactions respectively. Since $R_{\rm N}$ has opposite sign to $a$, a negative/positive curvature scalar indicates respective attractive/repulsive interactions. The absolute value of $R_{\rm N}$ provides a measure of the strength of these interactions.

Turning to the second term $(v-b)$, if $b=v$, all the space of the fluid system is occupied by the molecules, so no interactions are allowed. The corresponding curvature scalar vanishes, consistent with empirical observations of Ruppeiner geometry. If we do not impose a cutoff at the location of the well in the Lennard-Jones potential, shifting it to smaller $r$, then a short-range repulsive interaction is present. There is a range of negative $(v-b)$, and $R_{\rm N}$ becomes positive as expected.

Finally, considering the term $(ab+P v^3)$, it is positive since a fluid system has positive pressure and volume. But we have noted that when $T<\frac{a}{4b}$, the pressure becomes negative, yielding positive $R_{\rm N}$. This case also provides an opportunity for repulsive interactions, and is a combined effect of both $a$ and $b$.

The VdW parameter space has been shown to have a region where $R_{\rm N}$ is positive \cite{WeiWeiWei}, but it completely falls within the coexistence region. It was therefore excluded since it was not clear that the equation of state was applicable in the coexistence region. Only the attractive interaction then dominates in the whole parameter space, which is also consistent with the microscopic model of a VdW fluid. However if the equation of state is applicable in this region, the positive region of $R_{\rm N}$ is actually caused by negative pressure.

In summary, combining with the Lennard-Jones potential, we have given a clear microscopic interpretation of Ruppeiner geometry for a VdW fluid. We shall now apply this interpretation to black hole systems, constructing the corresponding molecular potentials.

\section{Schwarzschild AdS black hole}

\begin{figure}
\includegraphics[width=7cm]{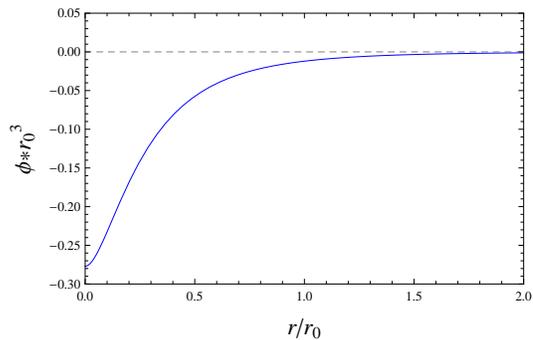}
\caption{The molecular potential (\ref{SchP}) for the Schwarzschild AdS black hole.}\label{PotSch}
\end{figure}

In an AdS space, Schwarzschild black hole is the most simplest black hole solution, which is described by the following action
\begin{equation}
 S=\frac{1}{16\pi G}\int d^4x\sqrt{-g} \left(R+\frac{6}{l^2}\right),
\end{equation}
where $l$ is the AdS radius and it is interpreted as the pressure of the black hole $P=3/(8\pi l^2)$ \cite{Kastor}. Solving the corresponding field equations, the line element reads
\begin{equation}
 ds^2=-f(r)dt^2+\frac{1}{f(r)}dr^2+r^2d\Omega^2_2,\label{metric}
\end{equation}
where the metric function $f(r)=1-2M/r+r^2/l^2$ and $M$ denotes the black hole mass. Requiring the absence of conical singularities at the horizon in the Euclidean section of the black hole solution, the Hawking temperature can be easily obtained
\begin{equation}
 T=\frac{\partial_rf(r_{\rm h})}{4\pi}=\frac{1}{4\pi r_{\rm h}}+2P r_{\rm h}\label{Hak}
\end{equation}
with $r_{\rm h}$ the radius of the black hole horizon, which is actually the equation of state of the black hole system. It is easy to check that the first law holds, i.e., $dE=TdS-PdV$, with the respective entropy and thermodynamic volume $S=\pi r_{\rm h}^2$ and $V=\frac{4}{3}\pi r_{\rm h}^3$. Noting that the specific volume $v=2r_{\rm h}l^2_{P}$ \cite{Kubiznak}, we have the number of the potential black hole molecules
\begin{eqnarray}
N=\frac{V}{v}=\frac{2}{3}\pi r_{\rm h}^2=\frac{2}{3}S,
\end{eqnarray}
from which we can write
\begin{eqnarray}
 P=\frac{NT}{V}-\frac{N^2}{2\pi V^2}=\frac{T}{v}-\frac{1}{2\pi v^2}
\end{eqnarray}
for the equation of state for the Schwarzschild AdS black hole. Comparing with the equation of state of the VdW fluid (\ref{eos}), we can immediately obtain
\begin{eqnarray}
 a=\frac{1}{2\pi}, \quad b=0.\label{abab}
\end{eqnarray}
After a simple calculation, we give a possible effective potential between two Schwarzschild AdS black hole molecules
\begin{eqnarray}
 \phi=\frac{495}{32 \sqrt{2}\pi^2 r_{0}^{3}}\left[\frac{1}{\left(r/r_{0}+\sqrt[6]{2}\right)^{12}}-\frac{
   1}{\left(r/r_{0}+\sqrt[6]{2}\right)^6}\right],
   \label{SchP}
\end{eqnarray}
which is plotted in Fig. \ref{PotSch}. Here the parameter $r_{0}$ has the dimension of length. Clearly, this gives an attractive interaction for $r/r_{0}\geq0$. At $r/r_{0}=0$, one has minimum $\phi_{\rm min}\approx-0.28/r_{0}^{3}$. This pattern is consistent with that of the normalized scalar curvature given in Ref. \cite{Mannw}, where only the attractive interaction dominates for the Schwarzschild AdS black hole.

\section{Charged AdS black hole}

Four-dimensional charged AdS black holes were the first ones found to possess a small-large black hole phase transition \cite{Kubiznak}, reminiscent of the liquid-gas phase transition of a VdW fluid. The line element is still of the form (\ref{metric}), but with metric function $f=1-2M/r+Q^2/r^2+r^2/l^2$, with $Q$ being the black hole charge. Compared with the Schwarzschild AdS black hole, the existence of the phase transition is mainly caused by the present of the black hole charge. The equation of state is
\begin{eqnarray}
 P=\frac{NT}{V}-\frac{N^2}{2\pi V^2}+\frac{2N^4Q^2}{\pi V^4}=\frac{T}{v}-\frac{1}{2\pi v^2}+\frac{2Q^2}{\pi v^4}
 .\label{eosbh}
\end{eqnarray}
A phase transition exists caused by the charge. Regarding the black hole system as being composed of underlying molecules \cite{Weiw,Weiwa2}, the charge must be an intrinsic property of these molecules. It is then natural to define an effective specific charge
\begin{eqnarray}
 q=N*\frac{Q}{V},
\end{eqnarray}
from which the equation of state (\ref{eosbh}) can be cast into a new form
\begin{eqnarray}
 P=\frac{T}{v}-\frac{1}{2\pi v^2}(1-4q^2).
\end{eqnarray}
Comparing with (\ref{eos}) of the VdW fluid, we obtain
\begin{eqnarray}
 a=\frac{1}{2\pi}(1-4q^2), \quad b=0.
\end{eqnarray}
When the black hole charge vanishes, $q$=0, reducing to (\ref{abab}) for the Schwarzschild AdS black hole case.

\begin{figure}
\includegraphics[width=7cm]{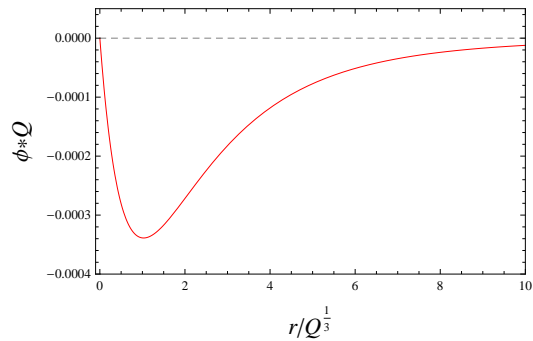}
\caption{The molecular potential (\ref{pch}) for the charged AdS black hole after fitting the local well with the zero point of the normalized curvature scalar \cite{Weiwa2,WeiWeiWei}.}\label{ChargedAdS}
\end{figure}

The charged AdS black hole microstructure can be dominated by repulsive interactions for small black holes \cite{Weiwa2}. Following the microscopic interpretation of Ruppeiner geometry, this can be understood in terms of a Lennard-Jones like potential that is not cut off at the well. A short-range repulsive interaction is present that will lead to a dominant repulsive interaction after a mean-field approximation. When the charge is included, one can construct the corresponding potential from (\ref{SchP}). However as we know, the molecular potential is not unique for the same equation of state. Here we would like to model the potential with a new interesting form
\begin{eqnarray}
 \phi=4\phi_{0}\left[\left(\frac{r_{0}}{r+r_{0}}\right)^{12}
  -\left(\frac{r_{0}}{r+r_{0}}\right)^{6}\right],
 \label{pch}
\end{eqnarray}
where the parameter $r_{0}$ is
\begin{eqnarray}
 r_{0}^{3}=\frac{495\left(1-4q^2\right)}{248\pi^2\phi_{0}},
\end{eqnarray}
which clearly depends on the black hole charge. For this case, the well is located at $r_{\rm min}=(2^{1/6}-1)r_{0}$. When $0<r<r_{\rm min}$, the slope of the potential is negative indicating the short-range repulsive interaction. Otherwise, the interaction is attractive. When $r=r_{\rm min}$, the interaction vanishes. If we expect it is related with that zero point of the normalized curvature scalar, we can fix the parameter $\phi_{0}$. From Refs. \cite{Weiwa2,WeiWeiWei}, we find that when $v=2Q(8\sqrt{2}\pi/3)^{1/3}$, the curvature scalar vanishes. After setting it to the volume $4\pi r_{\rm min}^3/3$ characterized by the well of the potential, we get
\begin{eqnarray}
 \phi_{0}\approx\frac{0.0003388}{Q}.
\end{eqnarray}
We describe the potential $\phi$ (\ref{pch}) for the charged AdS black hole in Fig. \ref{ChargedAdS}. Obviously, the short-range repulsive interaction is present at small $r/Q^{1/3}$. It is worth to point out that the small charge limit is $q\rightarrow 0$, or equivalently $r_{\rm h}^2Q\rightarrow 0$.

\section{Summary}

By combining empirical observations Ruppeiner geometry with the microscopic model of a VdW fluid with the Lennard-Jones potential, we have obtained the first microscopic interpretation for the constituent degrees of freedom of a (charged) AdS black hole. We have given for the first time explicit molecular potentials for these constituents, successfully implementing the potential black hole molecule model proposed previously \cite{Weiw,Weiwa2}. Our results address the outstanding problem of why attractive/repulsive interactions between the microscopic constituents of a thermodynamic system empirically corresponding negative/positive Ruppeiner curvature scalars.

Our results suggest that the repulsive interaction comes from two aspects. One is the finite size of the molecular constituents; the other is the cutoff of the well, where the short-range repulsive interaction should be present in the potential. In particular, we found that a repulsive interaction is present for the charged AdS black hole whilst absent for the Schwarzschild AdS black hole.
Note that the molecules have a finite actual size after an approximate comparison with the VdW fluid \cite{Miao}, whereas in our case, these black hole molecules are point particles that may have no actual size but have a specific volume $v=V/N$, similar to ideal gas molecules.

We close by noting that although our microscopic interpretation of Ruppeiner geometry was obtained via an AdS black hole, it is applicable to other asymptotically flat black holes due to the fact that the curvature scalar is invariant under the coordinate transformation in the thermodynamic parameter space \cite{Ruppeiner}. It would be interesting to calculate the molecular potentials for other black holes with more interesting black hole phase transition. Moreover, adopting these molecular potentials, it becomes possible to calculate correlation lengths, which are useful for studying critical phenomena in black hole phase transitions. These results should provide considerably more detailed information concerning the underlying degrees of freedom of black holes.

\section*{Acknowledgements}
The authors would like to thank Prof. Hong Zhao for useful discussions. As this paper was being completed, we became aware of Ref. \cite{Dutta}, in which the configuration integral was used to propose an interaction between the black hole molecules for a dyonic black hole. This work was supported by the National Natural Science Foundation of China (Grants No. 12075103, No. 11675064, No. 11875151, and No. 12047501), the 111 Project (Grant No. B20063), and the Natural Sciences and Engineering Research Council of Canada.

\end{document}